# Electrical/piezoresistive effects in bent Alpide MAPS

on behalf of the ALICE collaboration
M.J. Rossewij[1], E. M. Okkinga, H. M. Naqvi, R. G. E. Barthel and S. R. Alving

*Utrecht University,*
*Institute for Gravitational and Subatomic Physics (GRASP),*
*Princetonplein 1, 3584 CC, Utrecht, The Netherlands*
*E-mail:* m.j.rossewij@uu.nl

ABSTRACT: The ITS3 upgrade baseline design employs MAPS (Monolithic Active Pixel Sensor) in bent state. Bending experiments with the existing ITS2 MAPS (=Alpide chip) show it remains functional but with relative large analog supply current changes. It is shown that by the piezoresistive effect, rotation of current mirror FETs can be responsible which was confirmed after validating the layout. Measured Gauge Factor has proper sign but is 3 times lower than typical values derived from literature. The magnitude of the measured strain induced PMOS $V_{th}$ shift is as expected but the sign differs for compressive strain with some of the literature.

KEYWORDS: piezo resistive effect, threshold shift, stress, strain, bending, MAP, ALPIDE, ITS3

## Contents

---

[1] Corresponding author.

## 1. Introduction

The ITS3 design [1] foresees the replacement of the three innermost layers of the current ALICE tracker (ITS2) with wafer-scale stitched MAPS (65 nm) bent to cylindrical shapes. Multiple (beam) tests [2] with the existing ITS2 MAPS (=ALPIDE chip, 180 nm) showed that the ALPIDE retains functionality and performance when operated in a bent state.

However, compared to the bending induced strain ($\varepsilon=\Delta l/l \approx 0.1\%$), there was observed a relative large change ($\approx$+10% or -5% depending on bending axis) in analog power supply current ($I_a$). Gauge factors GF=$(\Delta R/R)/\varepsilon \approx -(\Delta I_a / I_{a0})/\varepsilon$ of -100 or +50 strongly deviate from e.g. metals which have GF$\approx$+2. Section 2 of this paper discusses how the electrical properties of semiconductors and MOSFETs are affected by applying stress/strain. Section 3 briefly introduces the ALPIDE chip and the setup for bending it. Section 4 shows the measured bending effects and the analysis on how these effects propagate though the chip and how the large $I_a$ changes can be explained.

## 2. Strain effects

A resistor ($R = \rho\, l/A$) of length l and cross-sectional area A made of a uniform material with resistivity $\rho$ and poisson's ratio $\nu$ changes under strain by:

$$\frac{\Delta R}{R} = \frac{1}{R}\left(\frac{\partial R}{\partial l}\Delta l + \frac{\partial R}{\partial A}\Delta A + \frac{\partial R}{\partial \rho}\Delta\rho\right) = \frac{\Delta l}{l} - \frac{\Delta A}{A} + \frac{\Delta\rho}{\rho} = (1+2\nu)\,\varepsilon\, + \frac{\Delta\rho}{\rho} \qquad (2\text{-}1)$$

For metals is eq. (2-1) dominated by geometrical changes. Therefore after neglecting the $\Delta\rho$ term, the Gauge Factor becomes GF=$(\Delta R/R)/\varepsilon \approx 1+2\nu$ which is typically around 2 and independent of crystal orientation. However in semi-conductors, $\Delta\rho$ becomes strongly dominant resulting in Gauge Factors up to 100 times larger (as for $I_a$) which can even have negative sign and are crystal orientation dependent. This gives GF=$1+2\nu+\pi E \approx \pi E$ with E being the Young's modulus and $\pi$ is the piezo resistive coefficient relating the resistance change with applied stress $\sigma$: $\Delta\rho/\rho=\pi\sigma=\pi E\varepsilon$.

### 2.1 Piezo resistive effect in homogenous semiconductors

Strain affects the lattice interatomic distances which in turn changes the semiconductor energy band structure. This affects the charge carrier effective mass m* which is proportional with the curvature of the fermi surface: $\left(\frac{1}{\boldsymbol{m}^*}\right)_{ij} = \frac{1}{\hbar^2}\frac{E(\boldsymbol{k})}{\partial k_i \partial k_j}$ \qquad (2-2)

Trough the mobility $\boldsymbol{\mu}=q\tau/\boldsymbol{m}^*$, $\boldsymbol{\rho}=(qn_e\boldsymbol{\mu_e} + qn_h\boldsymbol{\mu_h})^{-1}$ becomes anisotropic depending on the stress $\boldsymbol{\sigma}$ direction and crystal orientation as explained in Figure 1.

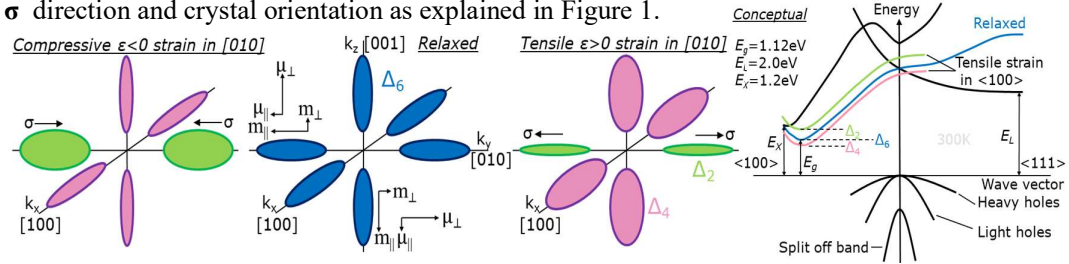

**Figure 1.** The 6 <100> energy valleys. The ellipsoidal shape makes **m*** direction dependent with $m_\perp$ (perpendicular to k) $\approx$ ⅕$m_\parallel$ ($m_\parallel \approx m_0$=electron mass). Although **m*** is direction dependent for each valley, the resistance $\rho$ becomes homogenous (with m*=$3/(m_\parallel^{-1}+m_\perp^{-1}+m_\perp^{-1})$ =0.26$m_0$) in unstrained silicon as all 6 degenerate ($\Delta_6$) valleys (blue) are equally populated. With e.g. uniaxial strain along [010], the 2 ($\pm$k) direction related energy levels ($\Delta_2$) differ from the 4 ($\Delta_4$) perpendicular. (Therefore $\pi_\perp \approx -\pi_\parallel/2$ for N-silicon) This affects directly **m*** through the fermi surface changes but also indirectly through the repopulation of the carriers over the 6 valleys following Boltzmann statistics. So due to the broken symmetry, the carriers are not equally populated in the 6 directions making the valley **m*** direction dependence become important.

## 2.2. Phenomenological description through the piezo resistance tensor π

In anisotropic material, π becomes a 4th rank tensor with in general $3^4$=81 independent coefficients relating the 2nd rank tensors of resistance change with stress: $\Delta\rho_{ij}/\rho = \pi_{ijkl}\sigma_{kl}$. This reduces significantly to include only three independent coefficients ($\pi_{11}$, $\pi_{12}$, $\pi_{44}$) due to the m3m symmetry of the silicon crystal [3]. The tensor is often written in six vector (Voight) notation as:

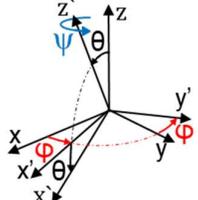

$$\frac{1}{\rho}\begin{bmatrix}\Delta\rho_1\\\Delta\rho_2\\\Delta\rho_3\\\Delta\rho_4\\\Delta\rho_5\\\Delta\rho_6\end{bmatrix} = \begin{bmatrix}\pi_{11}&\pi_{12}&\pi_{12}&0&0&0\\\pi_{12}&\pi_{11}&\pi_{12}&0&0&0\\\pi_{12}&\pi_{12}&\pi_{11}&0&0&0\\0&0&0&\pi_{44}&0&0\\0&0&0&0&\pi_{44}&0\\0&0&0&0&0&\pi_{44}\end{bmatrix}\begin{bmatrix}\sigma_1\\\sigma_2\\\sigma_3\\\sigma_4\\\sigma_5\\\sigma_6\end{bmatrix} \approx -\frac{1}{\mu}\begin{bmatrix}\Delta\mu_1\\\Delta\mu_2\\\Delta\mu_3\\\Delta\mu_4\\\Delta\mu_5\\\Delta\mu_6\end{bmatrix} \quad (2\text{-}3)$$

If applied stress **σ** is along one of the principal crystal axis <100>, eq. (2-3) can be used directly. Else transformations $T_{\alpha\beta}$ [4] using Euler angles (φ about z-axis, θ about y'axis and then ψ about z` axis again) relate the unprimed system referring to the crystal axis with new primed coordinate system where the piezo resistance becomes: $\pi'_{\alpha\beta} = T_{\alpha\gamma}\pi_{\gamma\delta}T^{-1}_{\delta\beta}$ (2-4)

The FETs in microchips are produced by patterning structures on/in a Si-wafer which nowadays mostly have (100) surface orientation (though (110) or (111) also exist). The microchips are normally diced in rectangular shape and the FETs are placed in Manhattan style [10] with the FET channels parallel to the dicing edges. The wafer scale ITS3 chip will also be diced in a rectangular shape (red box Figure 2) and have its bending axis parallel to a dicing axis, so **σ** will be aligned with a dicing axis.

In theory, the piezo resistive properties can be optimized by changing the angles from Figure 2 where both the uniaxial stress orientation φ (by rotating the wafer before patterning/dicing) and the resistor alignment λ (by having the FETs (R) under an angle on the mask) are referred to the [100] crystal axis. Using eq. (2-4) with ψ=θ=0 one obtains for e.g. (100) surface orientation [4]:

$\Delta R/R = \frac{1}{2}(\pi_{11} + \cos[2\lambda]\cos[2\varphi](\pi_{11} - \pi_{12}) + \pi_{12} + \sin[2\lambda]\sin[2\varphi]\pi_{44})\sigma_{x'}$ (2-5)

In practice, the dicing edge and therefore the stress orientation will be aligned with the wafer flat which is mostly in <110> direction. From eq. (2-5) $\pi_{\parallel}$=½($\pi_{11}$+$\pi_{12}$+$\pi_{44}$) with φ=λ=45° and $\pi_{\perp}$=½($\pi_{11}$+$\pi_{12}$-$\pi_{44}$) with φ=45° & λ=135°. Table 1 lists the piezo resistive coefficients with **σ** in <100> direction and the more common <110> direction. N silicon [6] in <100> has $\pi_{12}$≈-$\pi_{11}$/2 reflecting the ratio of the valleys and $\pi_{44}$≈0 as shearing stress affects all valleys similarly.

Table 1: piezoresistive coefficients ($10^{-12}$ Pa$^{-1}$) [7] for (100) surface orientation

| σ along | <100> | | | <110> | |
|---|---|---|---|---|---|
| | $\pi_{\parallel}$ | $\pi_{\perp}$ | $\pi_{44}$ | $\pi_{\parallel}$ | $\pi_{\perp}$ |
| | $\pi_{11}$ | $\pi_{12}$ | | ½($\pi_{11}$+$\pi_{12}$+$\pi_{44}$) | ½($\pi_{11}$+$\pi_{12}$-$\pi_{44}$) |
| N | -1022 | 534 | -136 | -312 | -176 |
| P | 66 | -11 | 1381 | 718 | -663 |
| NMOS | -426 | -207 | | -355 | -145 |
| PMOS | 91 | -62 | | 717 | -338 |
| E (GPa) | 130.2 | | | 168.9 | |
| ν | 0.279 | | | 0.064 | |

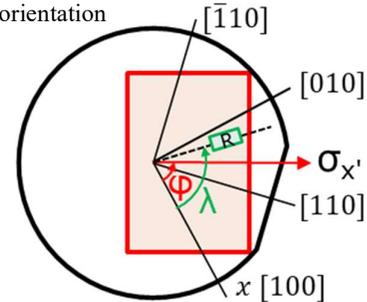

**Figure 2.** Rotating wafer and FETs

Using the numbers from Table 1 for N-silicon with **σ** along <100> gives GF$_{\parallel}$= E$\pi_{\parallel}$=-133 and GF$_{\perp}$= E$\pi_{\perp}$=+69 which shows some similarities with the GF mentioned in section 1 suggesting piezo resistance could indeed at the basis of the large $I_a$ changes. However a proper analyses should consider the piezo effects in FETs in stead of homogenous semiconductor.

## 2.3. Stress/strain effects in MOSFETs

As MOSFETs, operated in strong inversion (either linear or saturation region), have the current controlled by the channel resistive region, a similar piezoresistive response as homogenous silicon can be expected [8]. Modelling the response (e.g. using band structure calculation combined with the Monte Carlo simulation[9]) gets complicated by [10] scattering on the channel interface and the removal of the valley degeneration due to the vertical potential well. Therefore, the piezo resistive tensor with empirically measured coefficients (Table 1) are often used instead.

Strain also induces MOSFET threshold shifts $\Delta V_{th}$ which can be approximated [15] from changes in the silicon bandgap ($\Delta E_g = -3.78\varepsilon$ for $\varepsilon>0$ & $\Delta E_g = +6.19\varepsilon$ for $\varepsilon<0$) and the carrier mobility ($\Delta\mu$) as show in eq. (2-6). For pMOSFET is m=1.35 making the $-m\Delta E_g$ term is always positive:

$$q\Delta V_{th} \approx -m\Delta E_g + mk_B T ln\left(\frac{\mu_h(\varepsilon)}{\mu_h(0)}\right) \approx -m\Delta E_g + mk_B T ln(1 - \pi_h E\varepsilon), \quad m = 1.35 \quad (2\text{-}6)$$

## 3. Measurement Setup

### 3.1 ALPIDE MAPs pixel sensor and its central biasing circuitry

The bending experiments were performed on the ALPIDE chip [11] which are also in use in ITS2 [12]. This is a 0.5Mpixel MAPS chip where the FETs in each pixel are biased with voltages and currents coming from central DAC circuitry (Figure 4) containing following 3 (8-bit) DAC types:
1) 5 Current DACs: IBIAS, ITHR, IDB, IRESET, IAUX
2) 5 Voltage DACS with linear buffer: VCLIP, VCASP, VCASN, VCASN2, VTEMP
3) 4 Voltage DACs with Source Follower (SF): VRESETP, VRESETD, VPULSEH, VPULSEL

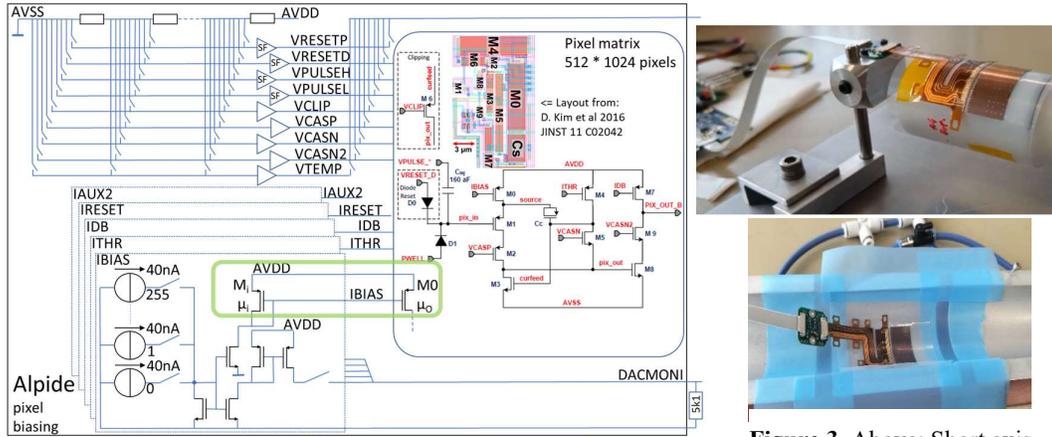

**Figure 4.** Alpide pixel bias circuitry. As DACs are central, all pixels share same biasing. All voltage DACs share same resistor divider. DACmonV/I pins allow direct monitoring of one of the voltage/current DACs

**Figure 3.** Above: Short axis convex ($\varepsilon>0$, tensile) bending. Below: Short axis concave ($\varepsilon<0$, compressive) bending.

### 3.2 Bending setup and DAQ

The measurement setup (Figure 3) allows convex and concave bending over long and short axis. The Alpide is bonded to a flex (modified for DACmonV/I access) and both are glued on a 50 μm thick plastic sheet which is bent over a mandril with curvature radii of 18, 24 or 30 mm as foreseen in ITS3. Concave bending uses porous aluminum profiles where vacuum keeps the Alpide-flex in place. The flex is connected through a standard FPC cable to a FPGA based DAQ board which can be controlled from a PC via USB. The DAQ board has ADCs to directly monitor DACmonV/I and the ALPIDE analog ($I_a$) and digital ($I_d$) supply current.

# 4. Results

## 4.1 Bending induced changes in ALPIDE analog ($I_a$) and digital ($I_d$) power supply current

Figure 5 shows measured $I_a$ and $I_d$ for different strain values derived from the bent-radii R using $\varepsilon=\frac{1}{2}t/R$ [10], with t=thickness of the ALPIDE. Bending has strong effect on $I_a$ while $I_d$ is mostly unaffected, especially compared to its absolute value which is $CV^2$ dominated where neither V nor C gets affected significantly by the bending. Blue lines in Figure 6 shows relative $I_a$ changes, giving GF=-83 for short axis. Long axis is asymmetric with GF=+52 ($\varepsilon>0$).

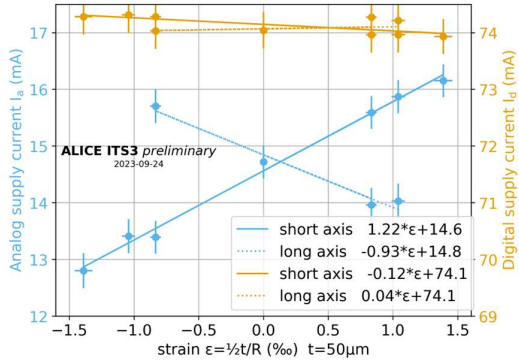 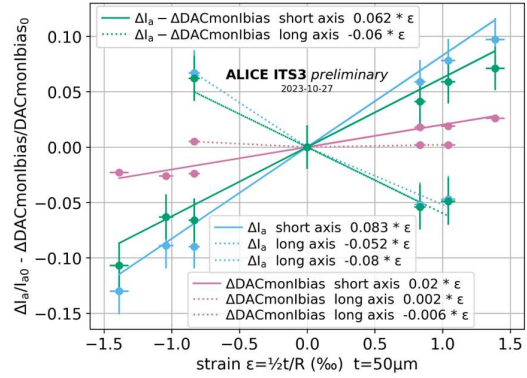

**Figure 5.** Measured $I_a$ and $I_d$ for different bent-radii       **Figure 6.** relative changes in $I_a$ - DACmonIbias

## 4.2 Bending induced changes in ALPIDE DACMONV/I

To investigate to which level the $I_a$ changes originate from changes in the central DAC outputs, DACmonV/I were measured for all bending radii. Figure 7 shows the difference of DACmonV/I in flat and bent (R=-30) state. It clearly shows the different responses from the 3 DAC types:

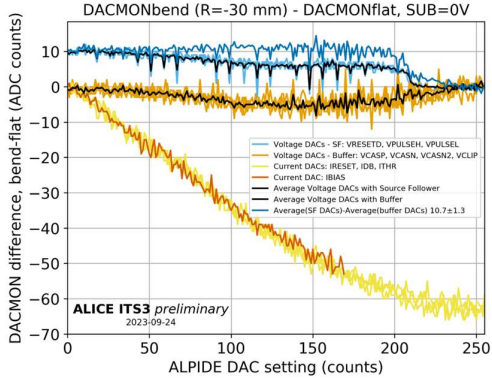 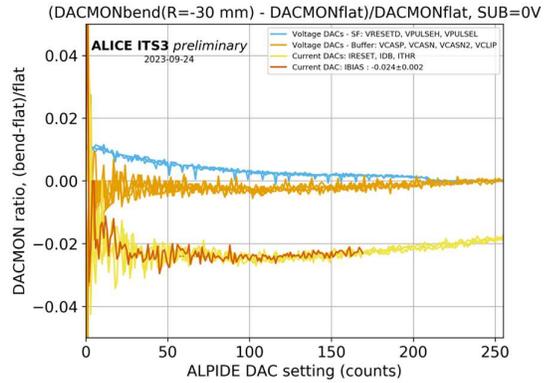

**Figure 7.** DACmonV/I difference for short axis       **Figure 8.** DACmonV/I ratio for short axis

1) The current DAC difference decreases linearly up to DAC setting 150. This translates in a -2% change in the current DACs as shown in figure 8 with DACmonV/I ratio i.s.o. difference.
2) All buffer voltage DACs show the same slight deviations which probably originate from the shared resistor divider.
3) The SF DAC voltage difference goes to 0 for DAC setting>200 as it is below SF PMOS threshold. For DAC setting <200, it seems rather straight, especially after correction with the averaged buffer voltage DACs values. This suggests it is also affected by the same errors from the shared resistor divider.

## 4.3 SF threshold shift

For all radii, the SF straight section is averaged over DAC setting 0-170, scaled (taking [9] into account that enhancement PMOS has a negative $V_{th}$, so a negative $\Delta V_{th}$ means an increase of $|V_{th}|$) to a $\Delta V_{th}$ in mV and plotted in Figure 9. It shows $|\Delta V_{th}|$ up to 7 mV at $\varepsilon \approx 1.4$‰ ($\hat{=}\sigma=E\varepsilon\approx236$ MPa) which corresponds to the reported (less then) 3mV/100MPa [11] and 5…20mV/GPa in [14] where figure 2 also shows similar trend. However in eq (2.6) becomes the $-m\Delta E_g$ term dominant for low strain (1.4‰<<5.6‰). So $\Delta V_{th}$ should always be positive which seems contradictive with [14] and the negative $\Delta V_{th}$ measured for $\varepsilon<0$. At present, the origin of the difference is unclear and requires further investigation.

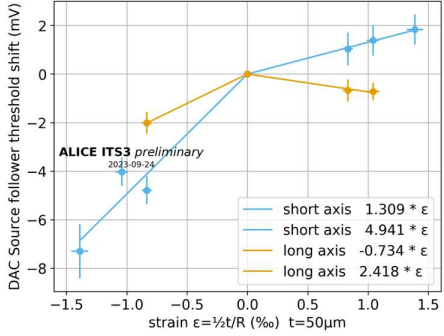
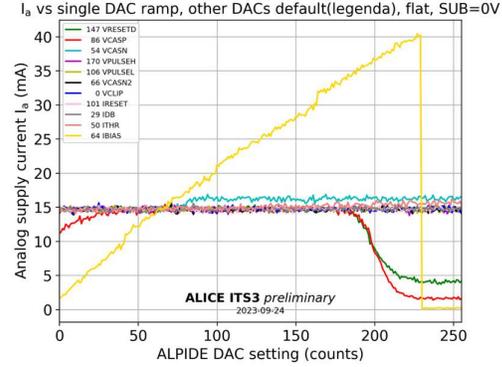

**Figure 9.** SF threshold shifts $\Delta V_{th}$ for different strain   **Figure 10.** $I_a$ measured for varying DAC settings

## 4.4 $I_a$ changes disentangled from IBIAS-DAC output changes

Figure 10 shows how $I_a$ changes when ramping up one DAC while keeping others at default. It shows $I_a$ mainly depends on Ibias in a linear way (until Ibias=235 where the DAQ board $I_a$ current protection gets activated). The purple curves in Figure 6 show the relative DACmonIbias changes at different bending radii. As mentioned before this is ±1…2%, much lower than the ±5...10% $I_a$ change. Subtracting DACmonIbias change from $\Delta I_a/I_{a0}$ gives the green curves in figure 6, basically representing part of the $I_a$ changes coming from the circuitry after DACmonIbias (=buffering + pixels). The long axis asymmetric behavior disappears and short axis gauge factor becomes smaller and similar to the long axis with opposite sign. Behavior like this is expected when current mirror FETs are rotated with each other [5]. Validating the design layout indeed showed that Ibias FET $M_0$ is rotated with respect to its counterpart $M_i$ (green box Figure 4).

The current mirror consist of 2 FETs ($M_i$ & $M_o$) with their sources and gates connected. So $V_{gs}$, $C_{ox}$, $V_{th}$ [8] are equal and from $I_D=\frac{1}{2}\mu C_{ox} W/L *(V_{gs}-V_{th})^2$ can be seen that $I_o = I_i \mu_o W_o L_i / \mu_i W_i L_o$. Designers can tune W and L to implement current multiplication factors. When $M_i$ & $M_o$ have different orientation and strain is applied, $\mu_o$ and $\mu_i$ will become different (strain will also affect W and L but as discussed in section 2 geometrical changes are small compared to changes in $\mu$). For (100) surface orientation and FETs perpendicular in <110> direction, stress along $M_i$ gives
$\mu_i' = \mu + \Delta\mu_i = \mu(1+\Delta\mu_i/\mu) = \mu(1-\Delta\rho_i/\rho) = \mu(1-\pi_\parallel\sigma)$ and $\mu_o' = \mu(1-\pi_\perp\sigma)$
$I_o' = I_o + \Delta I_o = I_o(1+\Delta I_o/I_o) = I_o \mu_o'/\mu_i' = I_o(1-\pi_\perp\sigma)/(1-\pi_\parallel\sigma) \approx I_o(1+(\pi_\parallel-\pi_\perp)\sigma) = I_o(1+\pi_{44}\sigma)$

So with stress along $M_i$, $\Delta I_a/I_a = (\pi_\parallel-\pi_\perp)\sigma = \pi_{44}\sigma$ and with stress along $M_o$, $\Delta I_a/I_a = (\pi_\perp-\pi_\parallel)\sigma = -\pi_{44}\sigma$. $M_o$ is vertical in the pixel and $M_i$ is horizontal in the periphery, so stress along $M_i$ occurs when the bending axis is parallel to the chip short axis. Then $\Delta I_a/I_a = \pi_{44}\sigma = \pi_{44}E\varepsilon = 1.055*169*\varepsilon = +0.18\varepsilon$(‰). The fit though the measurement is $+0.062\varepsilon$(‰), so the sign is ok but the magnitude is around 3 times lower. Explaining the difference requires further investigation considering e.g. differences (due to doping concentration) in $\pi$, $V_{th}$ shifts or other effects in the pixel circuitry.

## 5. Conclusion

Though the ALPIDE in bend state remains functional, $I_a$ shows large changes. It was showed that through the piezo resistive effect, rotation of current mirror FETs could be responsible which was confirmed after validating the layout. It emphasizes the importance of maintaining the orientation in symmetric structures especially when stress is applied. The magnitude of the measured PMOS $V_{th}$ shift is as expected but the sign differs for compressive strain with some of the literature.

## Acknowledgments

We would like to thank Franck Agnese and Christophe Wabnitz (CNRS-IPHC) for preparing the flexes with ALPIDEs that were used in the measurements.